\documentclass[twocolumn,preprintnumbers,amsmath,floatfix,prb,aps,superscriptaddress]{revtex4}

\usepackage{graphicx}
\makeatletter

\def\V{{V_{\rm pipe}}}

\begin{document} 

\newcommand{\be}{\begin{equation}}
\newcommand{\ee}{  \end{equation}}
\newcommand{\ba}{\begin{eqnarray}}
\newcommand{\ea}{  \end{eqnarray}}

\title{Microscopic model of a phononic refrigerator}

\author{Liliana Arrachea} \affiliation{Departamento de F\a'{i}sica,
  FCEyN and IFIBA, Universidad de Buenos Aires, Pabell\'on 1, Ciudad
  Universitaria, 1428 Buenos Aires, Argentina}

\author{Eduardo R. Mucciolo} \affiliation{Department of Physics,
  University of Central Florida, Orlando, Florida 32816, USA}

\author{Claudio Chamon} \affiliation{Department of Physics, Boston
  University, Boston, Massachusetts 02215, USA}

\author{Rodrigo B. Capaz} \affiliation{Instituto de F\a'{i}sica,
  Universidade Federal do Rio de Janeiro, 
Caixa Postal 68528, 
Rio de Janeiro
  21941-972, RJ, Brazil}

\date{\today}

%%%%%%%%%%%%%%%%%%%%%%%%%%%%%%%%%%%%%%%%%%%%%%%%%%%%%%%%%%%%%%%%%%%%
\begin{abstract}
  We analyze a simple microscopic model to pump heat from a cold to a
  hot reservoir in a nanomechanical system. The model consists of a
  one-dimensional chain of masses and springs coupled to a back gate
  through which a time-dependent perturbation is applied. The action
  of the gate creates a moving phononic barrier by locally pinning a
  mass. We solve the problem numerically using a non-equilibrium Green
  function technique. For low driving frequencies and for sharp
  traveling barriers, we show that this microscopic model realizes a
  phonon refrigerator.
\end{abstract}
%%%%%%%%%%%%%%%%%%%%%%%%%%%%%%%%%%%%%%%%%%%%%%%%%%%%%%%%%%%%%%%%%%%%%


\maketitle

%%%%%%%%%%%%%%%%%%%%%%%%%%%%%%%%%%%%%%%%%%%%%%%%%%%%%%%%%%%%%%%%%%%%%%
\section{Introduction}

In atomic gases, techniques such as evaporative cooling can bring
temperatures down to the submicrokelvin scale, allowing for the
observation of quantum phenomena such as Bose-Einstein
condensation. Progress in refrigerating condensed matter systems has
been less spectacular but steady; a recent example is the use of
active feedback for cooling nanomechanical cantilevered
beams.~\cite{cantilever} Other experimental examples, most of them in
electronic devices, have been reviewed in Refs. ~\onlinecite{elref}. On
the theoretical side, mechanisms for quantum refrigeration by means of
electronic pumping with ac fields operating at low \cite{liliheat} and
high \cite{kohler} frequencies have been recently proposed. There are
also several proposals for cooling nanomechanical systems by absorbing
phonons with electrons. \cite{el-phon}

The challenge of quantum refrigeration is to keep the heat extracted
from the cold sample higher than the energy dissipated into it by
operating the cooling machine. In electronic ac pumping, this
condition can be achieved within the so-called ``adiabatic''
quantum-pumping regime. For instance, in mesoscopic structures, this
regime is attained by modulating electronic potentials with two
low-frequency fields with a relative phase.\cite{liliheat} More
precisely, assuming that the electrons propagate coherently through
such a device, it can be shown that heat is pumped at a rate
proportional to $k_B T \Omega_0$, while the rate at which energy is
dissipated scales as $\hbar \Omega_0^2$, where $T$ is the temperature
of the cold reservoir and $\Omega_0$ is the frequency of the ac
field. As in the cases of charge \cite{chapump, adia} and spin
\cite{sppump} pumping, coherence is expected to be an important
ingredient for this type of heat pumping. These mechanisms
act on the electronic degrees of freedom, and therefore they are
ineffective in cooling insulators. Cooling procedures based on phonon
manipulation, however, apply equally well to metals and
insulators. Phononic refrigeration is the topic of this paper.
 
The minimal microscopic model for acoustic phonons consists of masses
coupled by springs. Heat transport and thermalization in chains of
coupled oscillators have a rather long history in statistical physics
since the pioneer work by Fermi, Pasta, and Ulam.\cite{fpu} Several
studies of different classical and quantum models have analyzed the
local temperature profile of chains of coupled oscillators in contact
with two reservoirs at different temperatures.\cite{modpho} The
striking feature found in these studies was the violation of the
Fourier law, according to which the local temperature is expected to
drop linearly along the chain. This behavior is the phononic
counterpart of the abrupt drop of the local voltage at the connections
between a finite size electronic system to two reservoirs at different
chemical potentials, which has been characterized by the concept of
``contact resistance''.\cite{cont-res} More recently, heat pumping has
been analyzed through a two-level system driven by an ac field coupled
to phononic baths at different temperatures,~\cite{segal} between
semi-infinite harmonic chains with a time-dependent
coupling,~\cite{li} and in anharmonic molecular junctions in contact
to phononic baths.\cite{moljun}

In a recent work,\cite{us} we have shown that when a barrier (or
perturbation) moves along a cavity (which we refer as pipe) connecting
two gases of acoustic phonons, a sizable amount of heat can be
transferred from the colder to the hotter gas. Assuming that the
barrier is perfectly reflective and its motion is slow enough to allow
for a fast thermalization of the phonon gas, we can imagine a cycle
where: (i) a barrier is inserted at the end of the pipe close to the
cold reservoir; (ii) the barrier is then adiabatically translated
through the pipe until it reaches the end in contact to the hot
reservoir; (iii) a second barrier is then inserted again near the cold
reservoir; and (iv) the first barrier is finally removed. The process
can then be repeated as a cycle. The result is a refrigerator which
does not rely on quantum coherence.

\begin{figure}
\includegraphics[width=\columnwidth,clip]{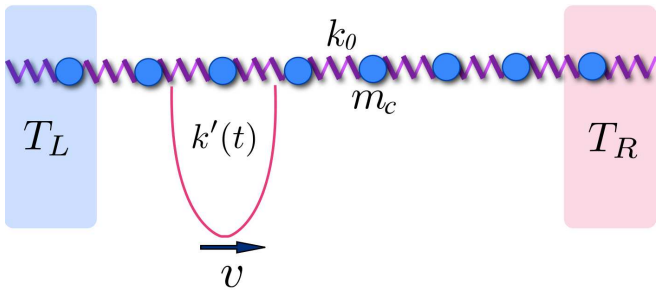}
\caption{\label{fig1} The one-dimensional microscopic model for
  acoustic phonons consists of a chain of particles (atoms of
  molecules) with equal mass $m_c$ coupled by springs with elastic
  constants $k_0$. The semi-infinite left portion of the chain is kept
  at a temperature $T_L$ which is lower than the temperature $T_R$ of
  the semi-infinite right portion. A traveling harmonic potential is
  applied locally to the central portion of the chain. This potential
  restricts the longitudinal motion of the particles -- see last term
  on the right-hand side of Eq. (\ref{hcent}). To perform a cooling
  cycle, the local potential is turned on and moved from near the cold
  to the hot reservoir at a constant speed $v$. When it reaches the
  hot reservoir, the potential is turned off and moved back to near
  the cold reservoir.}
\end{figure}

The phonon pump proposed in Ref.~\onlinecite{us} used a
thermodynamical formulation for the phonon gas. Here we present
instead a microscopic model of acoustic phonons that implements this
cooling scheme. The lattice model we consider consists of a series of
atoms or molecules with identical masses coupled by springs. The
moving barrier in this microscopic model is realized by modulating in
time and space a local pinning potential, which restricts the
longitudinal motion of the particles. The setup is shown schematically
in Fig. \ref{fig1}.

The paper is organized as follows. In Sec.~\ref{sec:theory} we present
the microscopic model. In Sec.~\ref{sec:results}, we show the result
of numerical computations of the heat currents and use it to evaluate
the lowest temperature the phonon pump can cool down. This result is
then compared to the prediction made in Ref.~\onlinecite{us} based on
a continuum model. In Sec.~\ref{sec:methods} we provide a detailed
description of the non-equilibrium Green function methodology used to
carry out the computations. Our conclusions are presented in
Sec.~\ref{sec:concl}. Some technical details are presented in the
Appendix.

%%%%%%%%%%%%%%%%%%%%%%%%%%%%%%%%%%%%%%%%%%%%%%%%%%%%%%%%%%%%%%%%%%%%%%
%%%%%%%%%%%%%%%%%%%%%%%%%%%%%%%%%%%%%%%%%%%%%%%%%%%%%%%%%%%%%%%%%%%%%%

%%%%%%%%%%%%%%%%%%%%%%%%%%%%%%%%%%%%%%%%%%%%%%%%%%%%%%%%%%%%%%%%%%%%%%
%%%%%%%%%%%%%%%%%%%%%%%%%%%%%%%%%%%%%%%%%%%%%%%%%%%%%%%%%%%%%%%%%%%%%%

%%%%%%%%%%%%%%%%%%%%%%%%%%%%%%%%%%%%%%%%%%%%%%%%%%%%%%%%%%%%%%%%%%%%%%
\section{Microscopic Model}
\label{sec:theory}

We consider the one-dimensional (1d) system sketched in
Fig. \ref{fig1}, with a central finite chain of $N$ atoms or molecules
(which we will refer as ``masses'' hereafter) with identical masses
$m_c$ coupled by springs of constant $k_c$. This system is connected
at its left (right) end to a semi-infinite chain $L$ ($R$) of masses
$m_L$ ($m_R$) coupled by springs of constant $k_L$ ($k_R$). These left
and right chains play the role of reservoirs, which we assume being
kept at temperatures $T_L$ and $T_R$, respectively. We only consider
longitudinal vibrational modes. The ensuing Hamiltonian reads
\begin{equation}
H=H_L+H_R+H_{c}(t)+H_{\rm cont}.
\end{equation}
The first two terms $H_{L,R}$ represent the reservoirs; the third
$H_c(t)$ describes the central chain, which serves as the pump and has
an explicit time dependence; the last term provides the contact
between the central chain $c$ and the $L,R$ reservoirs.

The masses in the central chain are coupled to their neighbors via
time-independent springs. An external potential is applied to the central
part of the chain in order to restrict (i.e., pin) the longitudinal
motion of the masses. We use a local harmonic approximation for this
pinning potential. (We defer to Sec. \ref{sec:concl} a discussion
about the possible experimental realizations.) The ensuing Hamiltonian reads
\begin{eqnarray}
\label{hcent}
H_{c}(t) & = & \sum_{l=1}^{N} \frac{p_{c,l}^2}{2 m_{c}} +
\sum_{l=1}^{N-1}\frac{k_0}{2} (x_{c,l}- x_{c,l+1} )^2 \nonumber\\ & &
+\ \sum_{l=1}^{N} \frac{k^{\prime}_{l}(t)}{2} \; x_{c,l}^2.
\end{eqnarray}
To illustrate the heat pumping mechanism, we break down the spring
constant of the local pinning potential into a train of pulses with a
Gaussian shape,
%~\cite{footnote1}
%
\begin{equation}
\label{bar} 
k^{\prime}_{l}(t) = \sum_{n=-\infty}^{+\infty} k^{(n)}_{l}(t),
\end{equation}
with $k^{(n)}_{l}(t) = k_1 \,e^{-\left[a l-v (t-n \tau)
    \right]^2/\sigma^2 }$, $n \tau \leq t \leq (n+1)\tau$, where $a$
is the lattice constant. The barrier speed is $v$, its width is
$\sigma$, and the time it takes to traverse the central chain is
$\tau= a N/v$. It is useful to define two characteristic frequencies
in the problem: the frequency $\omega_c = \sqrt{k_0/m_c}$, and the
pumping frequency $\Omega_0=2\pi/\tau$. (We shall work on units such
that $\hbar= k_B=1$.)

The contact between the central system and the reservoirs is described
by the Hamiltonian
\be
\label{hcont}
H_{\rm cont} = \frac{k_{L}}{2} (x_{L,1} - x_{c,1})^2 + \frac{k_R}{2}
(x_{R,1} - x_{c,N})^2.  \ee
Notice that the central region couples to the left reservoir at the
site $l=1$ and to the right one at $l=N$ (for later use, it is
convenient to define $l_L\equiv 1$, $l_R\equiv N$).

The reservoir Hamiltonians are given by
\begin{equation}\label{half}
H_{\alpha} = \sum_{j=1}^{N_{\alpha}}\left[ \frac{p_{\alpha,j}^2}{2
    m_{\alpha}} + \frac{k_{\alpha}}{2} (x_{\alpha,j} -
  x_{\alpha,j+1})^2 \right],
\end{equation}
with $\alpha = L,R$, in the limit of $N_{\alpha} \rightarrow \infty$
(which we shall take in due time). The lattice sites are labeled from
right to left for the $L$ reservoir and from left to right for the $R$
reservoir. It is convenient to express the degrees of freedom of the
$L$ and $R$ reservoirs in terms of normal modes. For open boundary
conditions, this corresponds to performing the following
transformation,
\begin{equation}
x_{\alpha, l} = \sqrt{ \frac{2}{ N_{\alpha}+1 } }
\sum_{n=0}^{N_{\alpha}} \sin (q^{\alpha}_n l) \; x_{\alpha, n}
\end{equation}
and
\begin{equation}
p_{\alpha, l} = \sqrt{ \frac{2}{N_{\alpha}+1}} \sum_{n=0}^{N_{\alpha}}
\sin (q^{\alpha}_n l) \; p_{\alpha, n},
\end{equation}
where
\be
\label{qal} 
q^{\alpha}_n = \frac{n\pi}
{N_{\alpha}+1},\;\;\;\;n=0,\ldots,N_{\alpha}.  \ee
The corresponding Hamiltonians transform into
\be H_{\alpha} = \sum_n \left\{\frac{p_{\alpha,n}^2}{2 m_{\alpha}} +
\frac{k_{\alpha}}{2} \left[ 1- \cos(q^{\alpha}_n) \right]
x_{\alpha,n}^2 \right\}.  \ee
%

%%%%%%%%%%%%%%%%%%%%%%%%%%%%%%%%%%%%%%%%%%%%%%%%%%%%%%%%%%%%%%%%%%%%%%%%%%
\subsection{Energy balance}
\label{sec:eb}

Following a procedure similar to that used in
Ref.~\onlinecite{liliheat}, we consider the variation in time of the
energy stored in one of the reservoirs. We find
\be
\label{cer}
\frac{dE_{\alpha}(t)}{dt} = J^Q_{\alpha}(t) + P_{l_{\alpha}}(t), \ee
where the first term is the heat current flowing from the connecting
site $l_\alpha$ of the central chain into the $\alpha=L,R$ reservoir,
which reads
\begin{equation}
\label{curalft}
J^Q_{\alpha}(t)= \sum_n k_{n,\alpha} \langle x_{\alpha,n}
\;\dot{x}_{l_{\alpha}} \rangle ,
\end{equation}
with $k_{n,\alpha}= k_{\alpha} \sqrt{ \frac{2}{ N_{\alpha}+1 } } \sin
(q^{\alpha}_n)$, while
\begin{equation}
\label{powla}
P_{l_{\alpha}}(t) = \frac{\partial} {\partial t} \,
k^{\prime}_{l_{\alpha}}(t) \,\langle x_{l_{\alpha}}^2 \rangle
\end{equation}
is the power due to the time-dependent forces acting at the connecting
site $l_{\alpha}$.

Conservation of energy implies that the {\em total} power invested by
the external fields is dissipated into the reservoirs at a rate

\be \label{cons} \overline{J}^Q_L + \overline{J}^Q_R= \sum_{l=1}^{N}
\overline{P}_{l}, \ee
where
\be \overline{P}_{l}= \frac{1}{\tau}\int_0^{\tau} dt\, P_l(t), \ee
is the power developed by the external fields at the site $l$ averaged
over one period, while
\be \label{dcj} \overline{J}^Q_{\alpha}= \frac{1}{\tau}\int_0^{\tau}
dt\, J^Q_{\alpha}(t), \ee
is the dc component of the heat current at the reservoir ${\alpha}$.

Notice that a net amount of work has to be invested in order to pump
heat from one reservoir to the other. Carrying out an analysis similar
to that developed in Ref. \onlinecite{liliheat} leads to the
conclusion that the rate at which the total work done by the external
fields is dissipated as heat flowing into the reservoirs is
proportional to $\Omega_0^2$.

In the case of heat transport induced by a temperature gradient
between the reservoirs, and in the absence of time-dependent
perturbations, the heat current (\ref{dcj}) can be in general
expressed as
\be \label{estalph}
J^{(0)}_{\alpha}=\sum_{\beta=L,R}\int_{-\infty}^{+\infty} \frac{d
  \omega}{2 \pi} \omega [n_{\beta}(\omega)-n_{\alpha}(\omega)] {\cal
  T}(\omega), \ee
where ${\cal T}(\omega)$ is the thermal transmission function of the
structure, while $n_{\alpha}(\omega)$ is the Bose-Einstein
distribution function, which depends of the temperature of the
reservoir $\alpha$.

We evaluate these physical quantities by employing the
non-equilibrium Green function formalism. Before describing the
technical details of the calculations, we first present the results
that demonstrate the refrigeration capabilities of the phonon pump.

%%%%%%%%%%%%%%%%%%%%%%%%%%%%%%%%%%%%%%%%%%%%%%%%%%%%%%%%%%%%%%%%%%%%%%%%
\section{Phononic refrigeration}

\subsection{Mechanism and minimum cooling temperature}
\label{sec:results}

We now show results for the heat currents corresponding to a
perturbation of the form of a train of traveling Gaussian pulses
described by the modulation shown in Eq. (\ref{bar}). The main goal
here is to compare the results obtained for our microscopic phononic
system to the predictions put forward in Ref. \onlinecite{us}, which
were derived for a continuum model system with a strong, delta-like
traveling perturbation. As mentioned in Sec. \ref{sec:eb} and in
Ref. \onlinecite{liliheat}, the rate at which heat is produced is
proportional to $\Omega_0^2$. Therefore, we consider a low driving
frequency $\Omega_0= 2 \pi/\tau$ in order to ensure a low rate of heat
dissipation into the reservoirs and to achieve a regime where heat can
be effectively transported from the cold to the hot reservoir. The
speed of the barrier is tuned to $v=a N/\tau$, such that a new pulse
appears on the left end of the central region every time a pulse moves
through the right end.

%%%%%%%%%%%%%%%%%%%%%%%%%%%%%%%%%%%%%%%%%%%%%%%%%%%%%%%%%%%%%%%%%%%%%%
%%%%%%%%%%%%%%%%%%%%%%%%%%%%%%%%%%%%%%%%%%%%%%%%%%%%%%%%%%%%%%%%%%%%%%
\begin{figure}
\includegraphics[width=0.9\columnwidth,clip]{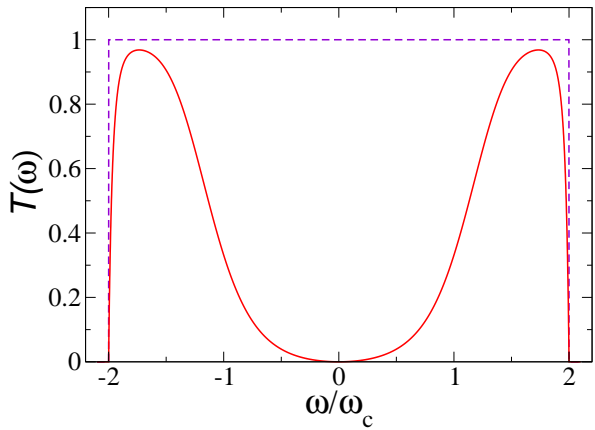}
\caption{The transmission function ${\cal T}(\omega)$ defined in
  Eq. (\ref{tran}) for a central system of $N=40$ atoms without any
  barrier (dashed lines) and with a Gaussian barrier with height $k_1
  = 1.5 k_0$ and $\sigma=a$ (solid lines) centered at $l=20$. The
  frequency is shown in units of the chain's characteristic frequency
  $\omega_c = \sqrt{k_0/m_c}$.}
\label{fig2}
\end{figure}
%%%%%%%%%%%%%%%%%%%%%%%%%%%%%%%%%%%%%%%%%%%%%%%%%%%%%%%%%%%%%%%%%%%%%%
%%%%%%%%%%%%%%%%%%%%%%%%%%%%%%%%%%%%%%%%%%%%%%%%%%%%%%%%%%%%%%%%%%%%%%

Before showing the results for the dc heat current, it is interesting
to analyze the effect of a stationary ($v=0$) pinning potential on the
behavior of the heat transmission function ${\cal T}(\omega)$ [see Eq.
  (\ref{tran})]. This is illustrated in Fig.~\ref{fig2}, where ${\cal
  T}(\omega)$ is shown for both the unperturbed chain ($k_1=0$) and
the case with a narrow, time-independent Gaussian spatial modulation
near the center of the chain. It is clear from Fig. \ref{fig2} that
the effect of the pinning potential is to decrease the transmission
probability in the low-energy sector of the phonon spectrum. Such an
effect is an indication that the pinning potential, when stationary at
a certain site, acts like a barrier, blocking heat transport from the
hot to the cold reservoir.

%%%%%%%%%%%%%%%%%%%%%%%%%%%%%%%%%%%%%%%%%%%%%%%%%%%%%%%%%%%%%%%%%%%%%%
%%%%%%%%%%%%%%%%%%%%%%%%%%%%%%%%%%%%%%%%%%%%%%%%%%%%%%%%%%%%%%%%%%%%%%
\begin{figure}
\includegraphics[width=0.9\columnwidth,clip]{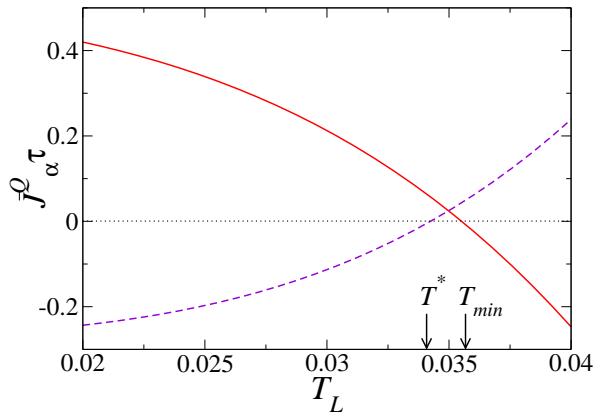}
\caption{(Color online) The average energy flowing through the
  contacts in a cycle, $\overline{J}^Q_{\alpha} \tau$, as functions of
  the temperature $T_L$ of the left (cold) reservoir (given in units
  of $\omega_c$). The temperature of the right reservoir is kept fixed
  $T_R=0.04 \omega_c$ and $\Omega_0=1\times 10^{-5} \omega_c$. The
  central system contains $N=80$ atoms. The barrier has a height
  $k_1=1.5 k_0$, $\sigma=a$, and moves from left to right with a speed
  $v= a N \Omega_0 /(2 \pi)$. The arrows on the horizontal axis
  indicate two temperature scales, $T^*$ and $T_{\rm min}$, defined in
  the main text.}
\label{fig3}
\end{figure}
%%%%%%%%%%%%%%%%%%%%%%%%%%%%%%%%%%%%%%%%%%%%%%%%%%%%%%%%%%%%%%%%%%%%%%
%%%%%%%%%%%%%%%%%%%%%%%%%%%%%%%%%%%%%%%%%%%%%%%%%%%%%%%%%%%%%%%%%%%%%%

In Ref. \onlinecite{us} we have shown that, besides blocking the usual
heat flowing in the direction of the heat gradient, a moving barrier
in a phonon gas also enables refrigeration under certain
conditions. Figure \ref{fig3} demonstrates that this is also the case
of the microscopic model considered in the present work. We consider a
finite-size central chain with $N=80$ atoms and a low pumping
frequency $\Omega_0 = 1 \times 10^{-5}\, \omega_c$. We fix the
temperature of the hot reservoir at $T_R=0.04\, \omega_c$ and analyze
the behavior of the heat current at the contacts between the central
system and each of the reservoirs. In Fig. \ref{fig3}, we can see
that, for large temperature differences $\Delta T=T_R-T_L$, the dc
heat current at the cold reservoir is positive, indicating that the
flow enters this reservoir, while it is negative at the hot one (the
flow exits that reservoir). In this situation, the direction of the
heat current is ruled by heat gradient (hot to cold). As $\Delta T$
decreases, there is a range of temperatures of the cold reservoir,
$T^* \leq T_L \leq T_{\rm min}$, where the two currents flowing into
the reservoirs become positive. When the temperature of the cold
reservoir overcomes $T_{\min}$, $\overline{J}^Q_L<0$ and
$\overline{J}^Q_R>0$. In this regime, there is a net heat flow from
the cold to the hot reservoir, which means that the system operates
like a refrigerator (i.e., heat flows against the temperature
gradient). It is also interesting to note that when cooling takes
place ($T>T_{\rm min}$), the absolute value of the currents
$\overline{J}^Q_{L,R}$ is approximately the same, indicating that the
dissipation of energy due to the action of the external forces is very
small [see Eq. (\ref{cons})].

%%%%%%%%%%%%%%%%%%%%%%%%%%%%%%%%%%%%%%%%%%%%%%%%%%%%%%%%%%%%%%%%%%%%%%
%%%%%%%%%%%%%%%%%%%%%%%%%%%%%%%%%%%%%%%%%%%%%%%%%%%%%%%%%%%%%%%%%%%%%%
\begin{figure}
\includegraphics[width=0.9\columnwidth,clip]{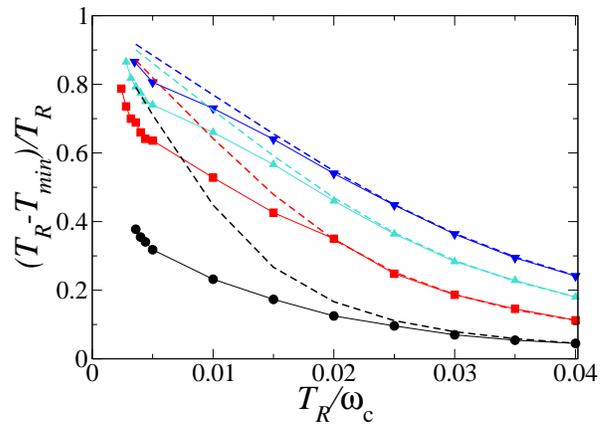}
\caption{(Color online) The relative temperature interval $(T_R-T_{\rm
    min})/T_R$ as a function of the right reservoir temperature$T_R$,
  where $T_{\rm min}$ is the minimum temperature of the left reservoir
  for which cooling is possible. Systems of sizes $N=40$, $80$, $120$,
  and $160$ corresponding to black circles, red squares, green down
  triangles, and blue up triangles, respectively, are considered (the
  solid lines are guides to the eye). For each system size, a fitting
  to Eq. (\ref{tmin}) is performed (dashed lines). Other parameters
  are the same as in Fig. \ref{fig3}.}
\label{fig4}
\end{figure}
%%%%%%%%%%%%%%%%%%%%%%%%%%%%%%%%%%%%%%%%%%%%%%%%%%%%%%%%%%%%%%%%%%%%%%
%%%%%%%%%%%%%%%%%%%%%%%%%%%%%%%%%%%%%%%%%%%%%%%%%%%%%%%%%%%%%%%%%%%%%%

In Ref. \onlinecite{us}, we have shown that an estimate for the
temperature $T_{\rm min}$ is given by
\be T_{\rm min} = \sqrt{\sqrt{\Theta_B^4 +T_R^4}-\Theta_B^2},
\label{tmin} 
\ee
where $\Theta_B = \lambda \sqrt{5 v/2\pi^3 c}$, with being $c$ the
sound velocity of the system and $\lambda$ being the amplitude of a
delta-function-like moving barrier [see Eq. (4) in
  Ref. \onlinecite{us}]. That calculation was done in a continuous
model assuming that the moving barrier produced negligible heat. It is
interesting to check if this prediction accurately describes cooling
in a microscopic model and still holds when there is a nonzero heat
generation production by the external fields. In order to build a
connection between the parameters used in these two distinct
formulations, we notice that the sound velocity in the microscopic
model is given by phonon velocity $c = a\sqrt{k_0/m_c}$, while the
delta-funtion amplitude can be related to the substrate spring
constant through $\lambda = (k_1/k_0)\, {\cal C}\, \omega_c$, where
${\cal C}$ is a geometrical factor related to the pulse shape.

Figure \ref{fig4} shows the dependence of the lowest cooling
temperature $T_{\rm min}$ on the hot reservoir temperature $T_R$ for
systems of several sizes and a common pumping frequency $\Omega_0 =1
\times 10^{-5}\, \omega_c$. Notice that fixing $\Omega_0$ requires
varying the barrier speed when considering central chains of different
sizes. In order to have a meaningful comparison for the different
chain sizes, we present the relative difference $(T_R-T_{\rm min}
)/T_R$. The dashed lines are obtained from Eq. (\ref{tmin}). The
geometrical factor ${\cal C}$ is used as a fitting parameter to adjust
Eq. (\ref{tmin}) to the numerical data. The fitted values for ${\cal
  C}$ are size dependent. We find ${\cal C} \sim 3.6, 4.1, 4.3, 4.5$,
for $N=40,80,120,160$, respectively. In all the plots,
Eq. (\ref{tmin}) provides a description of the numerical data for high
enough temperatures $T_R$. As the size of the central chain increases,
the range of temperatures where Eq. (\ref{tmin}) fits well the
numerical data systematically extends to lower values of $T_R$. The
discrepancy between Eq. (\ref{tmin}) and the numerical data for low
values $T_R$ can be understood by noticing that for low temperatures
the effect of heating due to the external fields becomes relatively
large. Indeed, the effect of heating is expected to be stronger for
the smaller systems, which is indeed where the discrepancy seems
larger. In summary, as the size of the system increases, the behavior
of the microscopic model approaches that of an infinite continuous
system and the effect of heating by the external field becomes
relatively insignificant. In the large-size limit, the results for the
microscopic model tend to coincide with the analytical prediction of
Eq. (\ref{tmin}) and this is precisely the observed behavior in
Fig. (\ref{fig4}).

%%%%%%%%%%%%%%%%%%%%%%%%%%%%%%%%%%%%%%%%%%%%%%%%%%%%%%%%%%%%%%%%%%%%%%%%%%%%%
\subsection{Cooling efficiency}
%\subsubsection{Phonon gas}

The results of Sec. \ref{sec:results} indicate that the model defined
in Sec. \ref{sec:theory} can be indeed regarded as a microscopic
realization of the phonon pump proposed in Ref.~\onlinecite{us}. Thus,
let us consider the efficiency of the phonon refrigerator in the
continuous, phonon gas model, and comparing it with that of a Carnot
cycle.

In Ref.~\onlinecite{us} we showed that the heat {\it extracted} from
the (cold) $L$ reservoir at the end of one cycle and the heat {\it
  injected} on the (hot) $R$ reservoir are given by
\begin{subequations}
\begin{eqnarray}
\label{eq:eL}
Q_L=-\Delta E_L^{L \to R} & = & \left(e_L+\frac{p_L}{\gamma\, d}
\right) \V ,\\
%\end{eqnarray}
%
%\begin{eqnarray}
\label{eq:eR}
Q_R=+\Delta E_R^{L \to R} & = & \left(e_L+\frac{p_R}{\gamma\, d}
\right) \V ,
\end{eqnarray}
where $\V$ is the volume swept by the moving barrier and $e_{L,R}$ and
$p_{L,R}$ are, respectively, the energy densities and the pressures in
both sides of the pump. The work done in the cycle is given by
\begin{equation}
\label{eq:W}
W= \frac{(p_R-p_L)}{\gamma\, d}\,\V \;.
\end{equation}
\end{subequations}
The pressure and energy of a phonon gas follow the relation $p =
-\left( \partial F/\partial V\right)_T = \gamma e$, where $\gamma$ is
the Gr\"uneisen parameter of the lattice and $d$ denotes the spatial
dimension \cite{AshcroftMermin,sorbello,lee}.

The cooling coefficient of performance of the refrigerator is given by
\begin{equation}
  \label{eq:cop}
  {\rm COP}=\frac{Q_L}{Q_R-Q_L}=(d+1)\,\frac{p_L}{p_R-p_L}\;.
\end{equation}
We can relate the pressures to the temperatures in the phonon gas in
$d$ dimensions and one finds (see Ref.~\cite{us}) that
$p_L/p_R=(T_L/T_R)^{d+1}$. We therefore arrive at the simple form
\begin{equation}
  \label{eq:cop2}
  {\rm COP}=(d+1)\,\frac{(T_L/T_R)^{d+1}}{1-(T_L/T_R)^{d+1}}\;.
\end{equation}
This cooling coefficient of performance is to be compared to that of a
Carnot cycle, ${\rm COP}_{\rm
  Carnot}=\frac{(T_L/T_R)}{1-(T_L/T_R)}$. It is useful to express the
ratio of these two coefficients of performance as
\begin{eqnarray}
  \label{eq:cop-ratio}
  \frac{{\rm COP}}{{\rm COP}_{\rm Carnot}} &=&
  \frac{(d+1)\,(T_L/T_R)^d}{1+(T_L/T_R)+\dots+(T_L/T_R)^d}
  \\ \nonumber\\ &\le& 1, \quad {\rm for}\;0\le T_L/T_R\le 1 \;.
  \nonumber
\end{eqnarray}
It is thus clear that the refrigerator is generically less efficient
than the Carnot one. When $T_L$ is close to $T_R$, the efficiency of
the phonon refrigerator approaches the Carnot limit. But when $T_L\ll
T_R$, and the efficiency is rather low compared to that of a Carnot
cycle. When $T_L/T_R=1/2$, ${\rm COP}/{\rm COP}_{\rm Carnot}$=2/3 for
one-dimensional pipes, and ${\rm COP}/{\rm COP}_{\rm Carnot}$=4/15 for
three-dimensional systems.

%%%%%%%%%%%%%%%%%%%%%%%%%%%%%%%%%%%%%%%%%%%%%%%%%%%%%%%%%%%%%%%%%%%%%%

%%%%%%%%%%%%%%%%%%%%%%%%%%%%%%%%%%%%%%%%%%%%%%%%%%%%%%%%%%%%%%%%%%%%%%
\section{Methodology: Green functions and Dyson equations}
\label{sec:methods}
%\subsection{Green functions and Dyson equations}

In order to develop an analytical framework where the nonequilibrium
heat currents in the microscopy phononic model can be evaluated, we
follow a procedure analogous to that introduced in
Ref. \onlinecite{liligreen} for fermionic systems. We begin by
defining the following lesser and greater Green functions,
\begin{equation}
D^<_{l,l'}(t,t') = i \langle x_{l'}(t') x_l(t) \rangle
\end{equation}
and
\begin{equation}
 D^>_{l,l'}(t,t') = i \langle x_{l}(t) x_{l'}(t') \rangle,
\end{equation}
as well as the retarded one,
\ba
\label{dret} 
D^R_{l,l'}(t,t') & = & -i\Theta(t-t')\langle [x_{l}(t), x_{l'}(t')]
\rangle \nonumber \\ & = & \Theta(t-t') [ D^<_{l,l'}(t,t') -
  D^>_{l,l'}(t,t') ].  \ea
For coordinates along the central chain (not including the contacts),
the Dyson equations for the retarded functions are
\ba & & -\left[ \partial^2_{t'} + \frac{2 k_0+ k^{\prime}_l(t)}{m_c}
  \right] D^R_{l,l'}(t,t') \nonumber \\ & & +\ \frac{k_0}{m_c}
D^R_{l,l'+1}(t,t')+ \frac{k_0}{m_c} D^R_{l,l'-1}(t,t') \nonumber \\ &
& = \frac{1}{m_c} \delta_{l,l'}\delta(t-t'), \ea
By writing the Dyson equation for $D^R$ along the contacts, it is
possible to integrate out the degrees of freedom of the reservoirs.
The result is
\ba & & -\partial^2_{t'} \hat{D}^R(t,t') - \hat{D}^R(t,t') \hat{M}(t')
- \nonumber \\ & & \int dt_ 1 \hat{\Sigma}^R(t,t_1) \hat{D}^R(t_1,t')
= \frac{1}{m_c} \delta_{l,l'} \delta(t-t'),
\label{dys} 
\ea
where $\hat{M}(t) = \hat{M}^{(0)} + \hat{M}^{(1)}(t)$, with
\begin{equation}
 M^{(0)}_{l,l'} = \left\{ \begin{array}{cc} \frac{k_0}{m_c} ( 2
   \delta_{l,l'} - \delta_{l',l \pm 1}), & 1<l<N, \\ (
   \frac{k_0}{m_c}+ \frac{k_L}{m_L}) \delta_{l,l'} - \frac{k_0}{m_c}
   \delta_{l',l + 1}, & l=1, \\ ( \frac{k_0}{m_c}+ \frac{k_R}{m_R})
   \delta_{l,l'} - \frac{k_0}{m_c} \delta_{l',l - 1}, &
   l=N, \end{array} \right.
\end{equation}
\begin{equation}
M^{(1)}_{l,l'}(t) = \frac{k^{\prime}_l(t)}{m_c} \delta_{l, l^{\prime}}
,
\end{equation}
\ba \Sigma^R_{l,l'}(t,t') &= & \sum_{\alpha=L,R} \delta_{l',l}\,
\delta_{l,l\alpha} \int_{-\infty}^{\infty}\frac{d\omega}{2 \pi} e^{-i
  \omega (t-t')}\nonumber \\ & & \times
\int_{-\infty}^{\infty}\frac{d\omega'}{2 \pi}
\frac{\Gamma_{\alpha}(\omega')}{\omega-\omega'+ i \eta}, \ea
and
\begin{eqnarray} 
\label{gamma}
\Gamma_{\alpha}(\omega)&=& \lim_{N_{\alpha} \rightarrow \infty} \frac{
  2 \pi (k_{ \alpha }/m_\alpha)^2 }{N_{\alpha}+1}
\sum_{n=0}^{N_{\alpha}}\sin^2(q^{\alpha}_n) \frac{1}{E_{\alpha,n}}
\nonumber \\ & & \times \left[ \delta(\omega-E_{\alpha,n}) -
  \delta(\omega+E_{\alpha,n}) \right] \nonumber \\ &= &
\mbox{sgn}(\omega) \frac{k_{\alpha} }{m_c} \Theta \Big( 1- \big(
\frac{ k_{\alpha}-m_{\alpha} \omega^2}{ k_{\alpha} } \big)^2 \Big)
\nonumber \\ & & \times \sqrt{ 1- \left(\frac{ k_{\alpha}-m_{\alpha}
    \omega^2}{ k_{\alpha} } \right)^2}.
\end{eqnarray}
Here, $\eta >0$ is an infinitesimal and $E_{\alpha,n} =
\sqrt{k_{\alpha}(1 - \cos q^{\alpha}_{n} )/m_{\alpha} }$.

On the other hand, the Dyson equation for the lesser Green functions
read
\ba
\label{dyless}
D^<_{l,l^{\prime}}(t,t^{\prime}) & = & \sum_{\alpha} \int dt_1 \int
dt_2\, D^R_{l,l_{\alpha}}(t,t_1)\, \Sigma_{\alpha}^<(t_1-t_2)
\nonumber \\ & & \times\ D^A_{l_{\alpha},l^{\prime} }(t_2,t^{\prime}),
\ea
and \ba
\label{dyless1}
 D^<_{l,q^{\alpha}_n}(t,t^{\prime}) & = &
 -\frac{k_{\alpha}}{m_{\alpha}} \int dt_1 [
   D^R_{l,l_{\alpha}}(t,t_1)\, d^<_{q^{\alpha}_n}(t_1-t^{\prime})
   \nonumber \\ & & +\ D^<_{l,l_{\alpha} }(t,t_1)\,
   d^A_{q^{\alpha}_n}(t_1-t^{\prime})], \ea
where Eq. (\ref{dyless}) corresponds to coordinates along the central
chain while Eq. (\ref{dyless1}) corresponds to one coordinate along
the chain and the other on the reservoir. Here we have introduced
\be \Sigma_{\alpha}^<(\omega) = i n_{\alpha}(\omega)
\Gamma_{\alpha}(\omega), \ee
\be d^A_{q^{\alpha}_n}(\omega) = \frac{1}{2 E_{\alpha,n}} \left[
  \frac{1}{\omega- i \eta - E_{\alpha,n} }- \frac{1}{\omega- i \eta +
    E_{\alpha,n}} \right], \ee
and
\be d^<_{q^{\alpha}}(\omega) = \frac{i \pi n_{\alpha}(\omega)}{2
  E_{\alpha,n}} \left[ \delta(\omega- E_{\alpha,n} )-\delta(\omega+
  E_{\alpha,n}) \right], \ee
with $n_{\alpha}(\omega)= 1/(e^{\omega/T_{\alpha}} - 1)$ being the
Bose-Einstein distribution, which depends on the temperature
$T_{\alpha}$ of the reservoir $\alpha$.

The strategy now is to solve Eq. (\ref{dys}) to evaluate the retarded
function and then use the result in Eqs. (\ref{dyless}) and
(\ref{dyless1}) to calculate the lesser functions.

%%%%%%%%%%%%%%%%%%%%%%%%%%%%%%%%%%%%%%%%%%%%%%%%%%%%%%%%%%%%%%%%%%%%%%
\subsection{Exact solution of the Dyson equation}
\label{sec:solution}

The exact functions $\hat{D}^R$ are obtained by solving the linear set
of equations represented by Eq. (\ref{dys}). It is also convenient to
use the representation
\be \hat{D}^R(t,t')=\sum_{k=-\infty}^{+\infty} e^{-i k \Omega_0 t}
\int_{-\infty}^{+\infty} \frac{ d \omega}{2 \pi} e^{-i \omega (t -t')}
\hat{\cal D}(k,\omega).
\label{fou} 
\ee
We get
\ba
\label{dyef1}
\hat{D}^R(t,\omega)& = & \hat{D}^{(0)}(\omega)+\sum_{k\neq 0, -\infty
}^{\infty} e^{-i k \Omega_0 t} \hat{D}^R(t,\omega + k \Omega_0)
\nonumber \\ & & \times\ \hat{M}^{(1)}_k \hat{D}^{(0)}(\omega), \ea
where we are considering the Fourier representation of the
time-dependent part of the Hamiltonian, $\hat{M}^{(1)}(t)= \sum_k
\hat{M}_k^{(1)} e^{- i k \Omega_0 t}$, while
\begin{equation}
\hat{D}^{(0)}(\omega)=[ \omega^2 \hat{1} - \hat{M}^{(0)} -
  \hat{\Sigma}^R(\omega)]^{-1}
\end{equation}
is the retarded Green function corresponding to the central chain of
harmonic oscillators connected to the reservoirs but free from the
time-dependent modulation of the spring constants.  In our case, the
frequency of the ac modulation is $\Omega_0= 2 \pi/\tau= 2 \pi N a/v$.
The exact solution for this set of coupled linear equations can be
obtained by following the procedure introduced in
Ref. \onlinecite{liligreen} for fermionic systems driven by ac
potentials. An alternative approach, appropriate for the case of a low
frequency $\Omega_0$ considered in the present work, is discussed in
Sec. \ref{adia}.

%%%%%%%%%%%%%%%%%%%%%%%%%%%%%%%%%%%%%%%%%%%%%%%%%%%%%%%%%%%%%%%%%%%%%%
\subsection{dc  heat currents}
\label{sec:dc}

We now turn to the evaluation of the dc component of the heat current
flowing from a given reservoir. To this end, we first express the heat
current in terms of Green functions. From Eq. (\ref{curalft}), the dc
component of the heat flow from the reservoir $\alpha$ can be
expressed as follows,
\be \overline{J}^Q_{\alpha} = \mbox{Re} \left\{\frac{1}{\tau}
\int_0^{\tau_0} dt \lim_{t^{\prime} \rightarrow t} \sum_n k_{n,\alpha}
\frac{-i \partial D^<_{l_{\alpha}, k_{n,\alpha} }(t,t^{\prime})}
     {\partial t^{\prime}} \right\}.  \ee
Using Eq. (\ref{dyless1}) and the representation shown in Eq.
(\ref{fou}), we get
\ba \overline{J}^Q_{\alpha} & = & \int \frac{d \omega}{2 \pi} \omega
\Big\{ 2 \mbox{Im}[{\cal D}_{l_{\alpha},l_{\alpha}}(0,\omega)]
\Gamma_{\alpha}(\omega) n_{\alpha}(\omega) \nonumber \\ & &
+\ \sum_{\beta} \sum_k |{\cal D}_{l_{\alpha},l_{\beta}}(k,\omega)|^2
\Gamma_{\beta}(\omega)\Gamma_{\alpha}(\omega + k \Omega_0)
n_{\beta}(\omega) \Big\}. \nonumber \\ \ea
The first term can be recast using Eq. (\ref{id1}), leading to the
following equation for the dc heat current,
\begin{eqnarray}
\label{dcalph}
\overline{J}^{Q}_{\alpha} & = &
\sum_{\beta=R,L}\sum_{k=-\infty}^{+\infty} \int_{-\infty}^{+\infty}
\frac{d \omega}{2 \pi} (\omega + k \Omega_0) [n_{\beta}(\omega)
  \nonumber \\ & & -\ n_{\alpha}(\omega+k\Omega_0)]
\Gamma_{\alpha}(\omega+k\Omega_0) \Gamma_{\beta}(\omega) |{\cal
  D}_{l_{\alpha},l_{\beta}}(k,\omega)|^2. \nonumber \\
\end{eqnarray} 
This equation indicates that a net heat current may exist even in the
absence of a temperature difference between the two reservoirs. Such a
flow contains in general an incoming component which accounts for the
work done by the ac fields and which is dissipated into the
reservoirs. As we discuss earlier, under certain conditions, a net
current between the reservoirs can be established; this current can go
against a temperature gradient, therefore allowing for cooling.

In the case of a temperature difference $\Delta T$ between the
reservoirs, $T_R=T_L+\Delta T$, and in stationary conditions, i.e. in
the absence of ac fields, Eq. (\ref{dcalph}) reduces to
Eq. (\ref{estalph})
%
%\begin{eqnarray}
%\label{estalph}
%J^{(0)}_{\alpha} & = & \sum_{\beta=R,L} \int_{-\infty}^{+\infty}
%\frac{d \omega}{2 \pi} \omega [n_{\beta}(\omega)-
 % n_{\alpha}(\omega)]\nonumber \\ & & \times\ \Gamma_{\alpha}(\omega)
%\Gamma_{\beta}(\omega) \left| D^{(0)}_{l_{\alpha},l_{\beta}}(\omega)
%\right|^2,
%\end{eqnarray}
%
where $J^{(0)}_{L}=-J^{(0)}_{R}=J^{(0)}$, being the thermal
transmission function
\be
\label{tran}
{\cal T}(\omega)= \Gamma_{\alpha}(\omega) \Gamma_{\beta}(\omega)
\left| D^{(0)}_{l_{\alpha},l_{\beta}}(\omega) \right|^2.  \ee
%

%%%%%%%%%%%%%%%%%%%%%%%%%%%%%%%%%%%%%%%%%%%%%%%%%%%%%%%%%%%%%%%%%%%%%%%%%%%%
\subsection{Low driving frequency solution of Dyson equation}
\label{adia}

Let us consider the Dyson equation in Eq. (\ref{dyef1}) in the limit
of low driving frequency $\Omega_0$.

%%%%%%%%%%%%%%%%%%%%%%%%%%%%%%%%%%%%%%%%%%%%%
\subsubsection*{1st order}

A solution exact up to ${\cal O}(\Omega_0)$ can be obtained by
expanding Eq. (\ref{dyef1}) as follows:
\ba \hat{D}(t,\omega) & \sim &\hat{D}^{(0)}(\omega) +
\hat{D}(t,\omega) \hat{M}^{(1)}(t) \hat{D}^{(0)}(\omega) + \nonumber
\\ & & i \partial_{\omega} \hat{D}(t,\omega) \frac{d
  \hat{M}^{(1)}(t)}{dt} \hat{D}^{(0)}(\omega).  \ea
We define the frozen Green function
\be \label{froz} \hat{D}_f (t,\omega) = \left[
  \hat{D}^{(0)}(\omega)^{-1} - \hat{M}^{(1)}(t) \right]^{-1}, \ee
in terms of which the exact solution of the Dyson equation at ${\cal
  O}(\Omega_0)$ reads
\be \hat{D}^{(1)}(t,\omega) = \hat{D}_f (t,\omega) + i
\partial_{\omega} \hat{D}_f (t,\omega) \frac{d \hat{M}^{(1)}(t)}{dt}
\hat{D}^{(0)}(\omega).  \ee
%

%%%%%%%%%%%%%%%%%%%%%%%%%%%%%%%%%%%%%%%%%%%%%%%%%%%%%
\subsubsection*{2nd order}

To obtain the solution exact up to ${\cal O}(\Omega_0^2)$, we consider
the following expansion of equation Eq. (\ref{dyef1}):
%
%\begin{widetext}
\ba \hat{D}(t,\omega) & \sim & \hat{D}^{(0)}(\omega) +
\hat{D}(t,\omega) \hat{M}^{(1)}(t) \hat{D}^{(0)}(\omega) \nonumber
\\ & & +\ i \partial_{\omega} \hat{D}(t,\omega) \frac{d
  \hat{M}^{(1)}(t)}{dt} \hat{D}^{(0)}(\omega) \nonumber \\ & &
-\ \frac{1}{2} \partial^2_{\omega} \hat{D}(t,\omega) \frac{d^2
  \hat{M}^{(1)}(t)}{dt^2} \hat{D}^{(0)}(\omega).  \ea
%\end{widetext}
%

The solution exact up to ${\cal O}(\Omega_0^2)$ is
\ba
\label{d2}
\hat{D}^{(2)}(t,\omega) & = & \hat{D}_f(\omega) + i \partial_{\omega}
\hat{D}^{(1)}(t,\omega) \frac{d \hat{M}^{(1)}(t)}{dt}
\hat{D}_f(\omega) \nonumber \\ & & -\frac{1}{2} \partial^2_{\omega}
\hat{D}_f (t,\omega) \frac{d^2 \hat{M}^{(1)}(t)}{dt^2}
\hat{D}_f(\omega) .  \ea
One can obtain this Green function numerically by first discretizing
the time in the interval $0 \leq t \leq \tau$, then solving
Eq. (\ref{d2}) for each one time, and finally evaluating the Fourier
transform to obtain the Floquet representation with which the dc
component of the heat current can be calculated from
Eq. (\ref{dcalph}).

%%%%%%%%%%%%%%%%%%%%%%%%%%%%%%%%%%%%%%%%%%%%%%%%%%%%%%%
\subsection{Low driving frequency expansion of the dc heat current}
\label{ljc}

Let us now use the low-frequency expansion in the expression of dc
heat current flowing into the reservoir $\alpha$ given by
Eq. (\ref{dcalph}). We first expand $n_{\alpha}(\omega + k \Omega_0)$
and $ \Gamma_{\alpha}(\omega)$ in powers of $\Omega_0$. By keeping
terms up to ${\cal O}(\Omega_0^2)$, we can identify the following
decomposition in the net current flowing into the reservoir \be
\overline{J}^Q_{\alpha}=\overline{J}^{th}_{\alpha}+\overline{J}^{p}_{\alpha}+
\overline{J}^{th-p}_{\alpha}+\overline{J}^{d}_{\alpha}, \ee where
``thermal'' component $\overline{J}^{th}_{\alpha} $ is due to the
gradient of temperature and flows from the hot to the cold reservoir,
the ``pumped'' component $\overline{J}^{p}_{\alpha}$ is purely induced
by the driving and accounts for the cooling mechanism, the ``mixed''
component $\overline{J}^{th-p}_{\alpha}$ results from an interference
process of the temperature gradient and the cooling mechanism, while
the ``dissipative'' term $\overline{J}^{d}_{\alpha}$ accounts for the
dissipation of energy into the reservoir due to the action of the ac
fields.

The explicit expression for the thermal component is
\ba \label{th} \overline{J}^{th}_{\alpha} &=& \int \frac{d \omega}{2
  \pi}[n_{\beta}(\omega)-n_{\alpha}(\omega)] \omega {\cal T}_f
(\omega), \nonumber \\ {\cal T}_f (\omega) &= &
\Gamma_{\alpha}(\omega) |{\cal D}_{f, l_{\alpha},l_{
    \beta}}(0,\omega)|^2 \Gamma_{\beta}(\omega), \ea
with $\beta \neq \alpha$, while ${\cal D}_{f, l_{\alpha},l_{
    \beta}}(0,\omega)$ is the dc component matrix element of the
frozen Green function (\ref{froz}). Notice that
$\overline{J}^{th}_L=-\overline{J}^{th}_R$. Also, recalling that the
left reservoir is the cold one, $\overline{J}^{th}_L>0$, indicating
that this component flows from right to left. It is also interesting
to note that the frozen transmission function corresponds to the chain
with the barrier. Thus, is practically vanishing for small $|\omega|$,
as shown in Fig. \ref{fig2}.

In order to write an expression for the pumped component, we take as a
reference temperature the one corresponding to the cold reservoir,
$T_L$, resulting
\ba \label{pump} \overline{J}^{p}_{\alpha} &=&\sum_{\beta=L,
  R}\sum_k\int \frac{d \omega}{2\pi} \omega k \Omega_0 \left( -\frac{d
  n_L(\omega) }{d \omega} \right) \nonumber \\ & & \times
\Gamma_{\alpha}(\omega) |{\cal D}_{f, l_{\alpha},l_{
    \beta}}(k,\omega)|^2 \Gamma_{\beta}(\omega).  \ea
In Appendix \ref{apb} we show that $\overline{J}^{p}_{L} =
-\overline{J}^{p}_{R}$. In our case, the driving mechanism is designed
to cool the left reservoir, thus, $\overline{J}^{p}_{L} <0$ and this
current flows against the temperature gradient. This component is
$\propto \Omega_0$ and we neglect eventual contributions $\propto \Omega_0^2$ or
terms of higher order in $\Omega_0$.

The mixed components read
\ba \label{mix} \overline{J}^{th-p}_L &=& \sum_{k} \int \frac{d
  \omega}{2 \pi}[n_{R}(\omega)-n_{L}(\omega)] \omega k \Omega_0 \times
\nonumber \\ & & \frac{\partial\Gamma_{L}(\omega) }{\partial
  \omega}|{\cal D}_{f, 1,N}(k,\omega)|^2 \Gamma_{R}(\omega),\nonumber
\\ \overline{J}^{th-p}_R&=& \overline{J}^{th-p}_{R,0}-
\overline{J}^{th-p}_L ,\nonumber \\ \overline{J}^{th-p}_{R,0}& = & -
\sum_{\beta=L,R}\sum_{k} \int \frac{d \omega}{2 \pi} \omega k \Omega_0
\frac{\partial^2 n_L (\omega) }{\partial \omega \partial T} \Delta T
\times \nonumber \\ & & \Gamma_{L}(\omega) |{\cal D}_{f,
  1,l_{\beta}}(k,\omega)|^2 \Gamma_{\beta}(\omega) \ea
which are $\propto \Omega_0 \Delta T $, being $\Delta T
=T_R-T_L$. Approximating the integrand of the last term, $\partial^2
n_L/(\partial \omega \partial T_L) \sim \partial n_L /(T_L\partial
\omega )$, which is the leading contribution in $\omega/T_L$, we get
$\overline{J}^{th-p}_{R,0} \sim \Delta T \overline{J}^p_R/T_L=-\Delta
T \overline{J}^p_L/T_L $.

Finally, in $\overline{J}_{\alpha}^{d}$, for which we do not provide
the explicit expression for sake of simplicity, we collect all the
terms that are $\propto \Omega_0^2$ and satisfy $\overline{
  J}_{\alpha}^{dis} >0, \;\alpha =L, R$. Thus, this contribution
describe pure dissipation of heat generated by the ac fields that
flows into the reservoirs. This contribution is vanishing small for
small $\Omega_0$.

%%%%%%%%%%%%%%%%%%%%%%%%%%%%%%%%%%%%%%%%%%%%%%%%%%%%%%%%%%%%%%%%%%%%
\section{Summary and Conclusions}
\label{sec:concl}

We have considered a microscopic model for the phononic refrigeration
mechanism proposed in Ref. \onlinecite{us}. The model describes a
one-dimensional lattice of atoms or molecules with identical masses
connected by springs and in contact with cold and hot reservoirs at
its ends. Considering only longitudinal vibrational modes, a moving
``barrier'' is realized by applying a local harmonic time-dependent
pinning potential to the masses. The cooling cycle consists of
propagating the barrier along the chain, going from the cold to the
hot reservoir with constant speed. When it reaches the hot reservoir,
the barrier is returned to near the cold reservoir and the cycle is
repeated. We have shown that such a perturbation, when static,
suppresses transmission within the low-energy sector of the phononic
spectrum. In this sense, our model describes a situation similar to
that found in electrostrictive polymer systems discussed in
Ref. \onlinecite{menezes10}, where an electric field is used to open a
local phonon gap.

The Hamiltonian described in Eq. (\ref{hcent}) could be realized 
experimentally in several ways. For instance, in the case of a chain
consisting of polarizable molecules, the local pinning of
intermolecular vibrations could be achieved through the application of
a inhomogeneous electric field, which would be swept along the chain
to produce a moving phonon barrier.

We note that the static version of the model described in
Sec. \ref{sec:theory} is rather standard in the study of heat
transport along  harmonic chains with a single mass per unit cell,
including terms beyond the usual intermass coupling but within the
harmonic approximation.\cite{modpho} In our case, we have considered a
time-dependent version of those extensions in order to model the
moving barrier.

The evaluation of heat currents requires the use of a formulation
capable of handling both a temperature gradient and the time-dependent
perturbation simultaneously. To that end, we employed an analytical
approach based on non-equilibrium Green functions. This treatment is
similar to the one introduced to study electronic systems driven
out-of-equilibrium by harmonic time-dependent voltages; here, it was 
suitably adapted to phonon propagators.\cite{liligreen} We have also
introduced a particular strategy that allows for solving Dyson
equations in the limit of weak driving frequencies and large
amplitudes. The propagators can then be evaluated numerically through
a time discretization and Fourier transform.

%%%%%%%%%%%%%%%%%%%%%%%%%%%%%%%%%%%%%%%%%%%%%%%%%%%%%%%%%%%%%%%%%%%%%%
\section{Acknowledgments}

This work is supported in part by ANPCyT and CONICET in Argentina, the
J. S. Guggenheim Memorial Foundation (LA), the DOE Grant
DE-FG02-06ER46316 (CC), and CAPES, CNPq, FAPERJ, and
INCT-Nanomateriais de Carbono in Brazil (RBC).

\appendix

%%%%%%%%%%%%%%%%%%%%%%%%%%%%%%%%%%%%%%%%%%%%%%%%%%%%%%%%%%%%%%%%%%%%%%
\section{An important identity}

Following the same steps of Ref. \onlinecite{liliheat}, we start from
the very definition of the retarded function in
Eq. (\ref{dret}). Then, using the representation of Eq. (\ref{fou}),
we obtain
\begin{widetext}
\begin{equation}
\int_{-\infty}^t dt' e^{i (\omega + i \eta) (t-t')} \Theta(t-t')
D^{<,>}_{l,l'}(t,t') = \sum_{k_1,k_2 } \sum_{\alpha=R,L}
\int_{-\infty}^{+\infty} \frac{ d\omega_1}{2 \pi}
\frac{D_{l,l_{\alpha}}(k_1,\omega_1) \Gamma_{\alpha}(\omega_1)
  D^*_{l',l_{\alpha}}(k_2,\omega_1)
  \lambda^{<,>}_{\alpha}(\omega)}{\omega + i \eta - (\omega_1 + k_2
  \Omega_0)},
\end{equation}
with $\lambda^{<}_{\alpha}(\omega) = -n_{\alpha}(\omega)$ and
$\lambda^{>}_{\alpha}(\omega)= -[1-n_{\alpha}(\omega)]$. Therefore,
\begin{equation}
{\cal D}_{l,l'}(k,\omega) = \sum_{k_1} \sum_{\alpha=R,L}
\int_{-\infty}^{+\infty} \frac{ d\omega_1}{2 \pi} \frac{ {\cal
    D}_{l,l_{\alpha}}(k_1,\omega_1) \Gamma_{\alpha}(\omega_1) {\cal
    D}^*_{l',l_{\alpha}}(k_2,\omega_1)}{\omega + i \eta - (\omega_1 +
  k_2 \Omega_0)},
\end{equation}
which implies the following identity:
\begin{equation}
\label{id1}
{\cal D}_{l,l'}(k,\omega)-{\cal D}_{l',l}^*(-k,\omega+k \Omega_0 ) =
-i \sum_{k^{\prime}},\sum_{\alpha=R,L} {\cal
  D}_{l,l_{\alpha}}(k+k^{\prime},\omega-k^{\prime} \Omega_0)
\Gamma_{\alpha}(\omega-k^{\prime} \Omega_0) {\cal
  D}^*_{l',l_{\alpha}}(k^{\prime},\omega-k^{\prime} \Omega_0).
\end{equation}
\end{widetext}

%%%%%%%%%%%%%%%%%%%%%%%%%%%%%%%%%%%%%%%%%%%%%%%%%%%%%%%%%%%%
\section{Conservation of the pumped component of the net heat current} 
\label{apb}

The goal of this appendix is to show that the pumped dc heat current
(\ref{pump}) is conserved, in the sense that it satisfies
\be \label{con} \sum_{\alpha=L,R }\overline{J}^{p}_{\alpha}= 0.  \ee
To this end, we use the relation between the frozen Green's funcion
and the frozen scattering matrix,\cite{liligreen}
\be S_{\alpha \beta}(k,\omega)= \delta_{\alpha,\beta}\delta_{k,0}- i
\sqrt{\Gamma_{\alpha}(\omega} {\cal D}_{f, l_{\alpha},l_{\beta}}(k,
\omega) \sqrt{\Gamma_{\beta}(\omega)}, \ee
and recast Eq. (\ref{pump}) as
\ba \overline{J}^p_{\alpha} &= &\frac{1}{ \tau}\int_0^{\tau} dt \int
\frac{d \omega}{2 \pi} \omega \left(-\frac{\partial
  n_L(\omega)}{\partial \omega} \right) \times \nonumber \\ & &
\mbox{Im} \left[ \hat{S}(t,\omega) \partial_t
  \hat{S}^{\dagger}(t,\omega)\right], \ea
which has been proved to satisfy Eq. (\ref{con}) on the basis of the
Birman-Krein relation, $d \mbox{ln}(det \hat{S}) = −\mbox{Tr}(\hat{S}
d\hat{S}^{\dagger} )$ (where $\mbox{det(}\hat{X} )$ and
$\mbox{Tr}(\hat{X} )$ are the determinant and the trace of a matrix
$\hat{X}$, respectively), applied to the frozen matrix, which is
unitary.\cite{adia,liliheat}

%%%%%%%%%%%%%%%%%%%%%%%%%%%%%%%%%%%%%%%%%%%%%%%%%%%%%%%%%%%%%%%%%%%%%%%%%%%%

%%%%%%%%%%%%%%%%%%%%%%%%%%%%%%%%%%%%%%%%%%%%%%%%%%%%%%%%%%%%%%%%%%%%%%%%%%%%

\end{document}